\documentclass[graybox]{svmult}

\usepackage{mathptmx}       
\usepackage{helvet}         
\usepackage{courier}        
\usepackage{type1cm}        
\usepackage{makeidx}         
\usepackage{graphicx}        
\usepackage{multicol}        
\usepackage[bottom]{footmisc}

\usepackage{amsmath,amssymb}

\begin{document}

\title*{Group analysis  of generalized  fifth-order Korteweg--de Vries equations  with time-dependent coefficients}
\titlerunning{Group analysis of  generalized fifth-order Korteweg--de Vries equations }
\author{Oksana Kuriksha, Severin Po\v{s}ta and Olena Vaneeva}
\institute{Oksana Kuriksha \at Petro Mohyla Black Sea State University, 10, 68 Desantnykiv Street, 54003 Mykolaiv, Ukraine. \email{oksana.kuriksha@gmail.com}
\and Severin Po\v{s}ta \at Department of Mathematics, Faculty of Nuclear Sciences and Physical Engineering,\\ Czech Technical University in Prague, 13 Trojanova Str., 120 00 Prague, Czech Republic.\\ \email{severin.posta@fjfi.cvut.cz} \and Olena Vaneeva \at Institute of Mathematics of the
National Academy of Sciences of Ukraine, 3 Tereshchenkivs'ka Street,
01601 Kyiv-4, Ukraine. \email{vaneeva@imath.kiev.ua}}
\maketitle

\abstract{We perform enhanced Lie symmetry analysis of generalized fifth-order Korteweg--de Vries equations with time-dependent coefficients. The corresponding similarity reductions are  classified and some exact solutions are constructed.}

\section{Introduction}

In this paper the class of  generalized variable-coefficient fifth-order Korteweg--de Vries (fKdV) equations
\begin{eqnarray}\label{eq_gg5kdv}
u_t+u^nu_x+\alpha(t)u+\beta(t)u_{xxxxx}=0
\end{eqnarray}
is investigated from the Lie symmetry point of view. Here $\alpha$ and $\beta$ are smooth nonvanishing functions
of the variable $t$ and $n$ is a positive integer, $n \geqslant 2$. This work is a natural continuation of the study undertaken by ourselves in~\cite{KPV_DUBNA}, where the group classification of the equations~(\ref{eq_gg5kdv}) with $n=1$ was carried out exhaustively. Lie symmetry analysis of the class~(\ref{eq_gg5kdv}) was initiated in~\cite{Wang&Liu&Zhang2013}. We  show that the results presented therein are incorrect. The case $n=2$ was considered also in~\cite{Wang2} but the complete group classification was not achieved.

Various  generalizations of the Korteweg--de Vries equation  appear  in many physical models, including ones describing
gravity waves, plasma waves and waves in lattices~\cite{jeffrey}. The equation (\ref{eq_gg5kdv}) with $n=1$, $\alpha=0$ and $\beta={\rm const}$ models, for example, one-dimensional hydromagnetic waves in a cold quasi-neutral collision-free plasma
propagating along the $x$-direction
under the presence of a uniform magnetic field under some conditions, namely,
when the propagation angle of the wave relative to the external magnetic
field becomes special, critical angle~\cite{Kakutani&Ono1969}. More references on studies concerned with these equations can be found in~\cite{KPV_DUBNA}.

The presence of variable coefficients in a  differential equation that model certain real-world phenomenon often allows one to get  better description of the phenomenon but, at the same time, makes the related studies of this equation, including group classification problems, more difficult.
In recent works on Lie symmetry analysis it was shown that the usage of admissible transformations in many cases is a cornerstone that leads to exhaustive solution of group classification problems~\cite{bihlo,Kuriksha&Posta&Vaneeva2014JPA,Popovych&Bihlo2012,popo2010a}.
That's why we firstly investigate admissible transformations in the class~(\ref{eq_gg5kdv}) in the next section and then proceed with the classification of Lie symmetries in Section~3. The corresponding reductions of equations~(\ref{eq_gg5kdv}) admitting extensions of Lie symmetry algebras are performed in Section~4, some exact solutions are constructed therein.  We discuss the incorrectnesses of the results obtained in~\cite{Wang&Liu&Zhang2013,Wang2} in the conclusion.

\section{Admissible transformations}
An admissible transformation (called also form-preserving~\cite{Kingston&Sophocleous1998} or allowed~\cite{Winternitz92} one)  can be regarded as a triple consisting  of two fixed equations from a class
and a~point transformation linking these equations~\cite{popo2010a}. The set of admissible  transformations of a class of differential equations naturally possesses the groupoid structure with respect to the standard operation of transformations composition~\cite{Popovych&Bihlo2012}. More details and examples on finding and usage of  admissible transformations for generalized fKdV equations as well as definitions of different kinds of equivalence groups can be found in~\cite{Kuriksha&Posta&Vaneeva2014JPA,VPS2013}.

We search for admissible transformations in class~(\ref{eq_gg5kdv}) using the direct method~\cite{Kingston&Sophocleous1998}, i.e., we
suppose that equation~(\ref{eq_gg5kdv}) is linked with an equation from the same class,
\begin{equation}\label{eq_gg5kdv_tilda}
\tilde u_{\tilde t}+\tilde u^{\,\tilde n}\tilde u_{\tilde x}+\tilde \alpha(\tilde t) \tilde u+\tilde \beta(\tilde t) \tilde u_{\tilde x\tilde x\tilde x\tilde x\tilde x}=0,
\end{equation}
by a nondegenerate point~transformation of the form
\begin{eqnarray}\label{EqEquivtransOfvcKdVlikeSuperclass}
\tilde t=T(t),\quad \tilde x=X^1(t)x+X^0(t),\quad \tilde u=U^1(t,x)u+U^0(t,x),
\end{eqnarray}
where $T$, $X^1$, $X^0$, $U^1$ and $U^0$ are arbitrary smooth functions of their variables with $T_tX^1U^1\neq0$.
We can restrict ourselves by consideration of point transformations of such a form instead of the most general form $\tilde t=T(t,x,u),$ $\tilde x=X(t,x,u),$ and $\tilde u=U(t,x,u)$, since the class~(\ref{eq_gg5kdv}) is a subclass (for $m=5$) of the more general class of evolution equations,
\begin{equation*}
u_t=F(t)u_m+G(t,x,u,u_1,\dots,u_{m-1}),
\end{equation*}
where
$
F\ne0, $ $G_{u_iu_{m-1}}=0,\ i=1,\dots,m-1,
$
and $m\geqslant 2$,
$
u_{m}=\frac{\partial^m u}{\partial x^m},
$
$F$ and $G$ are arbitrary smooth functions of their variables. It was proved in~\cite{VPS2013} that the latter class is normalized with respect to its equivalence group, where transformation components for independent and dependent variables are of the form~\eqref{EqEquivtransOfvcKdVlikeSuperclass}.

Now we perform the change of variables~(\ref{EqEquivtransOfvcKdVlikeSuperclass}) in equation~(\ref{eq_gg5kdv_tilda}).
The partial derivatives involved
in~(\ref{eq_gg5kdv}) are transformed as follows:
\begin{eqnarray*}\arraycolsep=0ex
&\tilde u_{\tilde t}  = \dfrac{1}{T_t}\left(U^1_tu+U^1u_t+U^0_t\right)-\dfrac{X^1_tx+X^0_t}{T_tX^1}\left(U^1_xu+U^1u_x+U^0_x\right),\\[1ex]
&\tilde u_{\tilde x}  =  \dfrac1{X^1}\left(U^1_xu+U^1u_x+U^0_x\right),\\[1ex]
&\tilde u_{\tilde x\tilde x\tilde x\tilde x\tilde x} = \dfrac{1}{(X^1)^5}\bigl(U^1_{xxxxx}u+5U^1_{xxxx}u_x+10U^1_{xxx}u_{xx}+10U^1_{xx}u_{xxx}+\\
&\qquad 5U^1_xu_{xxxx}+U^1u_{xxxxx}+U^0_{xxxxx}\bigr).
\end{eqnarray*}
We further substitute $u_t=-u^nu_x-\alpha(t)u-\beta(t)u_{xxxxx}$ to the obtained equation
in order to confine it to the manifold defined by~(\ref{eq_gg5kdv}) in the fifth-order jet space
with the independent variables $(t,x)$ and the dependent variable~$u$.
Splitting the obtained identity with respect to the derivatives of $u$ leads to the determining
equations on the functions~$T$, $X^1$, $X^0$, $U^1$ and $U^0$. Solving them we get,
in particular, the  conditions  \[\tilde n=n,\quad U^0=U^1_x=0,\quad \tilde \beta T_t-\beta(X^1)^5=0.\]
Then the rest of the determining equations result in
\begin{equation*}
X^1_t=X^0_t=0, \quad(U^1)^n T_t=X^1,\quad \tilde\alpha U^1T_t=\alpha U^1-U^1_t.
\end{equation*}
We solve these equations and get the the following assertion.


\begin{theorem}The generalized equivalence group~$G^{\sim}$ of the class~(\ref{eq_gg5kdv}) consists of the transformations
\begin{eqnarray*}
&\tilde t=T(t),\quad \tilde x=\delta_1x+\delta_2,\quad
\tilde u=\left(\dfrac{\delta_1}{T_t}\right)^\frac{1}{n} u, \\
&\tilde \alpha(\tilde t)=\dfrac{\alpha}{T_t}+\dfrac{T_{tt}}{n T_t^2}, \quad\tilde\beta(\tilde
t)=\dfrac{{\delta_1}^5}{T_t}\beta(t),\quad\tilde n=n,
\end{eqnarray*}
where  $\delta_j,$ $j=1,2,$ are arbitrary constants, $T$ is an arbitrary smooth function with $\delta_1T_t>0$.

The entire set of admissible transformations of the class~(\ref{eq_gg5kdv}) is
generated by the transformations from the group~$G^{\sim}$.
\end{theorem}
\noindent
{\bf Remark~1.}
 If we assume that the constant~$n$  varies in the class~(\ref{eq_gg5kdv}), then the equivalence group~$G^{\sim}$  is
generalized since $n$ is involved explicitly in the  transformation of the variable $u$.
Since $n$ is invariant under the action of transformations from the equivalence group,
the class~(\ref{eq_gg5kdv}) can be considered as the union of  its disjoint subclasses with fixed~$n$.
For each such subclass the equivalence  group $G^{\sim}$ is
usual one.

\medskip
Using Theorem~1 we derive a criterion of reducibility of variable-coefficient equations~(\ref{eq_gg5kdv}) to constant coefficient equations from the same class.
\begin{theorem}
A variable coefficient equation from the class~(\ref{eq_gg5kdv}) is reducible to the constant coefficient  equation from the same class if and only if its coefficients $\alpha$ and $\beta$ satisfy the equality
\begin{equation}\label{criterion1}
n\left(\alpha/\beta\right)_t=\left(1/\beta\right)_{tt}.
\end{equation}
\end{theorem}

Equivalence
transformations from the group~$G^\sim$  allow us to gauge one of
the arbitrary element $\alpha$ or $\beta$ to a simple constant value, for example, $\alpha$ can be set to zero or $\beta$ to unity.
The gauge $\alpha=0$ leads to more essential simplification of the study than the gauge $\beta=1$, therefore, the first one is preferable. Any equation from the
class~(\ref{eq_gg5kdv}) can be mapped to an equation from the same class with $\tilde
\alpha=0$
by the equivalence
transformation
\begin{equation}\label{eq_gauge}
\tilde t=\int\!e^{-n \int\!\alpha(t)\,{\rm d}t}{\rm d}t, \quad
\tilde x=x,\quad\tilde u=e^{\int\! \alpha(t)\, {\rm d}t}u.
\end{equation}  Then the single variable coefficient in the transformed equation will be expressed via $\alpha$ and $\beta$ as $\tilde\beta=e^{n\int\! \alpha(t)\, dt}\beta.$ (Here and in what follows an integral with respect to~$t$ should be interpreted as a fixed antiderivative.) Therefore, we can restrict
ourselves to the study of the class
\begin{eqnarray}\label{eq_g5kdv}
u_t+u^nu_x+\beta(t)u_{xxxxx}=0.
\end{eqnarray}
This will not lead to a loss of generality as all results on symmetries,
classical solutions and other related
objects for equations~(\ref{eq_gg5kdv}) can be constructed
using the similar results obtained for equations from the class~(\ref{eq_g5kdv}) and equivalence transformation~\eqref{eq_gauge}.

To derive the equivalence group for~(\ref{eq_g5kdv}) we set $\tilde\alpha=\alpha=0$ in the corresponding transformation
presented in Theorem~1 and deduce that the function $T$ is linear with respect to $t$. The following assertion is true.
\begin{corollary}The generalized equivalence group~$G^{\sim}_0$ of the class~(\ref{eq_g5kdv}) comprises  the transformations
\begin{equation}\label{tr_equiv_cor2}
\tilde t=\delta_3 t+\delta_4,\quad \tilde x=\delta_1x+\delta_2,\quad
\tilde u=\left(\frac{\delta_1}{\delta_3}\right)^\frac{1}{n} u, \quad
\tilde\beta(\tilde t)=\dfrac{{\delta_1}^5}{\delta_3}\beta(t),\quad\tilde n=n,
\end{equation}
where  $\delta_j,$ $j=1,2,3,4,$ are arbitrary constants with
$\delta_1\delta_3>0$.

The entire set of admissible transformations of the class~(\ref{eq_g5kdv}) is
generated by the transformations from the group~$G^{\sim}_0$.
\end{corollary}
Remark~1 is also true for the equivalence group~$G^{\sim}_0$.

\section{Lie symmetries}
In the previous section we have shown that the group classification problem for the class~(\ref{eq_gg5kdv}) reduces to the similar problem for its subclass~(\ref{eq_g5kdv}).
In order to carry out the group classification of~(\ref{eq_g5kdv})
we use the classical algorithm~\cite{Olver1986}.
Namely, we
look for symmetry generators of the form $Q=\tau(t,x,u)\partial_t+\xi(t,x,u)\partial_x+\eta(t,x,u)\partial_u$
and require that
\begin{equation}\label{c2}
Q^{(5)}\{u_t+u^nu_x+\beta(t)u_{xxxxx}\}=0
\end{equation}
identically, modulo equation~(\ref{eq_g5kdv}). Here  $Q^{(5)}$ is the fifth prolongation of the operator~$Q$~\cite{Olver1986,Ovsiannikov1982}.
Note that the restriction on $n$ to be integer is inessential for the group classification problem, so we can assume that $n$ is a real nonzero constant.

The infinitesimal invariance criterion implies
\[
\tau=\tau(t),\quad
\xi=\xi(t,x), \quad
\eta=\eta^1(t,x)u+\eta^0(t,x),
\]
where $\tau$, $\xi$, $\eta^1$ and $\eta^0$ are arbitrary smooth functions of their variables.
The rest of the determining equations have the form
\begin{eqnarray*}
&\tau \beta_t =(5\xi_x-\tau_t)\beta,\quad \eta^1_x=2\xi_{xx},\quad \eta^1_{xx}=\xi_{xxx},\quad 2\eta^1_{xxx}=\xi_{xxxx},\\
&\eta^1_xu^{n+1}+\eta^0_xu^{n}+(\eta^1_t+\eta^1_{xxxxx}\beta)u+\eta^0_t+\eta^0_{xxxxx}\beta=0,\\
&(\tau_t-\xi_x+n\eta^1)u^n+n\eta^0u^{n-1}+(5\eta^1_{xxxx}-\xi_{xxxxx})\beta-\xi_t=0.
\end{eqnarray*}
The derived determining equations were verified using GeM software package~\cite{Cheviakov}.
The latter two equations can be split with respect to different powers of $u$. Special cases of the splitting arise if $n=0$ or $n=1$.
If $n=0$ equations~(\ref{eq_g5kdv}) are linear ones and, therefore, excluded from the consideration.  The case $n=1$ is thoroughly investigated in~\cite{KPV_DUBNA}. So, we concentrate our attention on the case  $n\neq0,1$.

If $n\neq1$ the determining equations result in
\[\tau=(c_1-c_2n)t+c_3,\quad\xi=c_1x+c_0,\quad\eta^1=c_2,\quad\eta^0=0,\]
where $c_i$, $i=0,\dots,3$, are arbitrary constants. Thus, the infinitesimal generator has the form
\[
Q=((c_1-c_2n)t+c_3)\partial_t+(c_1x+c_0)\partial_x+c_2u\partial_u.
\]
The classifying equation on $\beta$ is
\begin{equation}\label{c11}
((c_1-c_2n)t+c_3)\beta_t=(4c_1+nc_2)\beta.
\end{equation}
To derive
the kernel~$A^{\rm ker}$ of maximal Lie invariance algebras~$A^{\rm max}$ of equations from the class~(\ref{eq_g5kdv}) (i.e., the Lie invariance algebra admitted by any equation from~(\ref{eq_g5kdv})) we split in~(\ref{c11}) with respect to $\beta$  and $\beta_t$.
Then $c_1=c_2=c_3=0$ and  $Q=c_0\partial_x$. Thus, $A^{\rm ker}=\langle\partial_x\rangle$.
To get possible extensions of $A^{\rm ker}$ we consider~(\ref{c11}) not as an identity but as an equation  on $\beta$, that has the form
 \begin{equation}\label{c12}
(pt+q)\beta_t=r\beta.
\end{equation}
The group classification of class~(\ref{eq_g5kdv}) is equivalent to the integration of the latter equation
up to the $G^\sim_0$-equivalence.
The equivalence transformations~(\ref{tr_equiv_cor2}) act on the coefficients $p,$ $q$, and $r$ of equation~(\ref{c12}) as follows:
\[
\tilde p= \kappa p,\quad \tilde q=\kappa(q\delta_3-p\delta_4),\quad\tilde r=\kappa r,
\]
where $\kappa$ is a nonzero constant. Therefore, there are three inequiva\-lent nonzero triples $(p,q,r)$:
$(1,0,\rho)$, $(0,1,1)$ and $(0,1,0)$, where $\rho$ is an arbitrary constant. We integrate~(\ref{c12}) for these values of $(p,q,r)$.
Up to $G^{\sim}_0$-equivalence $\beta$ takes the values from the set $\left\{\varepsilon  t^\rho,\,\varepsilon  e^t,\,\varepsilon\right\}$.
 Here $\rho$ and $\varepsilon$
are arbitrary constants with $\rho\varepsilon\neq0$, $\varepsilon=\pm1\bmod G^{\sim}_0$. The last step is to substitute the obtained forms of $\beta$ into
equation~\eqref{c11} and to find the corresponding values of $c_i,$ $i=0,\dots,3,$ that define the infinitesimal operator $Q.$
 We get that
all $G^\sim_0$-inequivalent cases of Lie symmetry extension are exhausted by the following:

\medskip

$\beta=\varepsilon  t^\rho,\ \rho\neq0\colon$
$Q=\frac5{\rho+1}c_1t\partial_t+(c_1x+c_0)\partial_x+\frac{\rho-4}{n(\rho+1)}c_1u\partial_u$,

\medskip

$\beta=\varepsilon  e^t\colon$
$Q=5c_1\partial_t+(c_1x+c_0)\partial_x+\frac1nc_1u\partial_u$,

\medskip
$\beta=\varepsilon\colon$
$Q=(5c_1t+c_3)\partial_t+(c_1x+c_0)\partial_x-\frac4nc_1u\partial_u$,

\medskip
\noindent
where $c_0$, $c_1$ and $c_3$ are arbitrary constants.
We have proved the following statement.
\begin{theorem}
The kernel of the maximal Lie invariance algebras of nonlinear equations from the class~(\ref{eq_g5kdv}) with $n\neq1$
coincides with the one-dimensional algebra $\langle\partial_x\rangle$.
All possible $G^\sim_0$-inequivalent  cases of extension of the maximal Lie invariance algebras are exhausted
by those presented in Cases 2--4 of Table~1.
\end{theorem}
\begin{table}[t!] \renewcommand{\arraystretch}{2}
\begin{center}
\textbf{Table~1.}
The group classification of the class~$u_t+u^nu_x+\beta(t) u_{xxxxx}=0$.
\\[2ex]
\begin{tabular}{|@{\,\,\,}c@{\,\,\,}|@{\,\,\,}c@{\,\,\,}|@{\,\,\,}l@{\,\,\,}|}
\hline
no.&$\beta(t)$&\hfil Basis of $A^{\max}$ \\
\hline
1&$\forall$&$\partial_x$\\
\hline
2&
$\varepsilon t^\rho$&$\partial_x,\,\,5nt\partial_t+(\rho+1)n x\partial_x+(\rho-4) u\partial_u$\\
\hline
3&$\varepsilon e^{t}$&
$\partial_x,\,\,5n\partial_t+nx\partial_x+u\partial_u$\\
\hline
4&$\varepsilon $&
$\partial_x,\,\,\partial_t,\,\,5nt\partial_t+nx\partial_x-4u\partial_u$\\
\hline
\end{tabular}
\\[2ex]
\parbox{110mm}{Here $\alpha=0\bmod\, G^\sim_0$,  $\rho$  is an arbitrary nonzero constant; $\varepsilon=\pm1\bmod\, G^\sim_0.$}
\end{center}
\end{table}
\begin{proposition}
A group classification list for the class~(\ref{eq_gg5kdv}) up to $G^\sim$-equivalence coincides with the list presented in Table~1.
\end{proposition}
\begin{proposition}
An equation of the form~(\ref{eq_gg5kdv}) admits a three-dimensional Lie symmetry algebra if and only if it is point-equivalent to the constant-coefficient fKdV equation $u_t+u^nu_x+\varepsilon u_{xxxxx}=0$ from the same class.
\end{proposition}
For convenience of further applications we present in Table~2
the complete list of Lie symmetry extensions for the initial class~(\ref{eq_gg5kdv}),
where arbitrary elements are not simplified by equivalence transformations (the detailed procedure of deriving such a list from a simplified one is described in~\cite{Vaneeva2012}).
\begin{table}[t!]\footnotesize \renewcommand{\arraystretch}{2}
\begin{center}
\textbf{Table~2.}
The group classification of the class~(\ref{eq_gg5kdv}) with $n\neq0,1$ using no equivalence.
\\[2ex]
\begin{tabular}{|@{\,\,\,}c@{\,\,\,}|@{\,\,\,}c@{\,\,\,}|@{\,\,\,}l@{\,\,\,}|}
\hline
no.&$\beta(t)$&\hfil Basis of $A^{\max}$ \\
\hline
1&$\forall$&$\partial_x$\\
\hline
2 & $\lambda T_t\big(T+\kappa\big)^{\rho}$&$\partial_x,\,\,5n(T+\kappa)T_t^{-1}\partial_t+n(\rho+1)x\partial_x+(\rho-4-5n\alpha(t)(T+\kappa)T_t^{-1}) u\partial_u$\\
\hline
3&$\lambda T_t  e^{m T}$&
$\partial_x,\,\,5nT_t^{-1}\partial_t+mn x\partial_x+\big(m-5n\alpha(t)T_t^{-1}\big)u\partial_u$\\
\hline
4&$\lambda T_t$& $\partial_x,\,\,T_t^{-1}(\partial_t-\alpha(t)u\partial_u),\,\,5nTT_t^{-1}\partial_t+nx\partial_x-(4+5n\alpha(t)TT_t^{-1}) u\partial_u$\\
\hline
\end{tabular}
\\[2ex]
\parbox{110mm}{Here $\lambda $, $\kappa$, $\rho$, and $m$ are arbitrary constants with $\lambda \rho m\neq0$, $T=T(t)=\int\!e^{-n \int\!\alpha(t)\,{\rm d}t}{\rm d}t$,  and
the function $\alpha(t)$ is arbitrary in all cases. }
\end{center}
\end{table}

The obtained group classification results  give  all equations~(\ref{eq_gg5kdv}) for which
the classical method of Lie reduction can be applied.

\section{Symmetry reductions and construction of exact solutions}
One of the most efficient techniques for construction of solutions for nonlinear partial differential equations is the Lie reduction method, based on the usage of Lie symmetries that correspond to Lie groups of continuous point transformations~\cite{Olver1986,Ovsiannikov1982}.
Any ($1{+}1$)-dimensional partial differential equation admitting a one-parameter Lie symmetry group (acting regularly and transversally on a manifold defined by this equation) can  be reduced to an ordinary differential equation.
Lie reduction method is well known and algorithmic~\cite{Olver1986,Ovsiannikov1982}.
In order to get an optimal system of group-invariant solutions reductions should be performed with respect to subalgebras from the optimal system~\cite[Section~3.3]{Olver1986}.

To find optimal systems of one-dimensional subalgebras for Lie algebras $A^{\rm max}$ presented in Table~1, we firstly consider their structure, using notations of~\cite{vaneeva:pate1977a}.
In Cases~2 and~3  the maximal Lie-invariance algebras are two-dimensional. In  Case 2 with $\rho=-1$  it is Abelian ($2A_1$). The algebras adduced in Case 2 with $\rho\neq-1$ and Case 3 are non-Abelian ($A_2$). The three-dimensional algebra with basis operators presented in Case~4 is of the type $A_{3.5}^a$, where $a=1/5.$

Therefore, optimal systems of one-dimensional subalgebras of the maximal Lie invariance algebras $A^{\rm max}$  presented in Table 1 are the following:

\smallskip
$2_{\rho\neq-1}\colon$\
${\mathfrak g}^{\,}_0=\langle\partial_x\rangle,
$\ \  ${\mathfrak g}^{\,}_{2.1}=\langle 5nt\partial_t+(\rho+1)n x\partial_x+(\rho-4) u\partial_u\rangle$;

\smallskip
$2_{\rho=-1}\colon$\
${\mathfrak g}^{\,}_0=\langle\partial_x\rangle,
$\ \ $
{\mathfrak g}^a_{2.2}=\langle nt\partial_t+a\partial_x-u\partial_u\rangle$,\ \
where $a$ is an arbitrary constant;

\smallskip
$3\colon$\
${\mathfrak g}^{\,}_0=\langle\partial_x\rangle,
$\ \  $
{\mathfrak g}^{\,}_3=\langle5n\partial_t+nx\partial_x+u\partial_u\rangle$;

\smallskip
$4\colon$\
${\mathfrak g}^{\,}_0=\langle\partial_x\rangle,
$\ \  $
{\mathfrak g}^\sigma_{4.1}=\langle\partial_t+\sigma\partial_x\rangle,$\ \
$
{\mathfrak g}^{\,}_{4.2}=\langle5nt\partial_t+nx\partial_x-4u\partial_u\rangle$;\ 
 $\sigma\in\{-1,0,1\}$.

\smallskip

We do not perform the reductions with respect to the subalgebra ${\mathfrak g}^{\,}_0$ since they lead to constant solutions only. The reductions with respect  to other one-dimensional subalgebras from the found optimal lists are presented in Table~3.

\begin{table}[h!] \renewcommand{\arraystretch}{2}
\begin{center}
\textbf{Table 3.} Similarity reductions of the equations~$u_t+u^nu_x+\beta(t)u_{xxxxx}=0$.
\\[2ex]
\begin{tabular}
{|@{\,\,\,}c@{\,\,\,}|@{\,\,\,}c@{\,\,\,}|@{\,\,\,}c@{\,\,\,}|@{\,\,\,}c@{\,\,\,}|@{\,\,\,}c@{\,\,\,}|@{\,\,\,}l@{\,\,\,}|}
\hline
no.&$\beta(t)$&$\mathfrak g$& $\omega$ &\hfil Ansatz&\hfil Reduced ODE
\\
\hline
1&$\varepsilon t^\rho,\ \rho\neq-1$&${\mathfrak g}^{\,}_{2.1}$&$xt^{-\frac{\rho+1}{5}}$ &
$u=t^{\frac{\rho-4}{5n}}\varphi(\omega)$ & $\varepsilon\varphi'''''+ \left(\varphi^n-\frac{\rho+1}{5}\omega\right)\varphi'+\frac{\rho-4}{5n}\varphi=0$
\\
\hline
2 &$\varepsilon t^{-1}$& ${\mathfrak g}^a_{2.2}$&$x-\frac{a}{n}\ln t$ & $u=t^{-\frac{1}{n}}\varphi(\omega)$ & $\varepsilon\varphi'''''+ \left(\varphi^n-\frac{a}{n}\right)\varphi'-\frac1n\varphi=0$
\\
\hline
3 &$\varepsilon e^t$&${\mathfrak g}_{3}$& $xe^{-\frac15t}$ & $u=e^{\frac1{5n}t}\varphi(\omega)$ & $
\varepsilon\varphi'''''+\left(\varphi^n-\frac1{5}{\omega}\right)\varphi'+\frac1{5n}\varphi=0$
\\
\hline
4 &$\varepsilon  $&${\mathfrak g}^\sigma_{4.1}$ &$x-\sigma t$ & $u=\varphi(\omega)$ & $
\varepsilon\varphi'''''+\left(\varphi^n-\sigma\right)\varphi'=0$
\\
\hline
5 &$\varepsilon  $&${\mathfrak g}_{4.2}$ &$xt^{-\frac15}$ & $u=t^{-\frac4{5n}}\varphi(\omega)$ & $
\varepsilon\varphi'''''+\left(\varphi^n-\frac\omega5\right)\varphi'-\frac4{5n}\varphi=0$
\\
\hline
\end{tabular}
\end{center}
\parbox{110mm}{Here $a$ is an arbitrary constant, $\sigma\in\{-1,0,1\}$, $\varepsilon=\pm1\bmod G^\sim_0,$ $n\neq0,1$.}
\end{table}
It is possible to consider also reductions of the generalized fKdV equations to algebraic equations using two-dimensional subalgebras of their Lie invariance algebras. There is only one such subalgebra that leads to a nonconstant solution, it is the subalgebra \[\langle\partial_t,\,\, 5nt\partial_t+nx\partial_x-4u\partial_u\rangle\] of the algebra $A^{\rm max}$ presented in Case~4 of Table~1. The corresponding ansatz $u=Cx^{-\frac4n}$
reduces the equation
\begin{equation}\label{5thKdV}
u_t+u^nu_x+\varepsilon u_{xxxxx}=0
\end{equation}
to an algebraic equation on the constant $C$. 
We solve it and get the stationary solution
\begin{equation*}
u=(-8\varepsilon(n+1)(n+2)(n+4)(3n+4))^\frac1n(nx)^{-\frac4n}.
\end{equation*}
of the equation~\eqref{5thKdV}. 
Using this solution and equivalence transformation~\eqref{eq_gauge} we construct simple  nonstationary exact solution,
\begin{equation*}
u=(-8\varepsilon(n+1)(n+2)(n+4)(3n+4))^\frac1n(nx)^{-\frac4n}e^{-\int\! \alpha(t)\, {\rm d}t},
\end{equation*}
 for the fKdV equation with time-dependent coefficients
\begin{equation}\label{5thKdVext}
u_t+u^nu_x+\alpha(t) u+\varepsilon e^{-n\int\! \alpha(t)\, dt} u_{xxxxx}=0,
\end{equation}
where $\alpha$ is an arbitrary nonvanishing smooth function.

If $n=2$ the travelling wave solution
\begin{equation*}
u=\pm2\sqrt{-10\varepsilon}\left(3\tanh(x+24\varepsilon t)^2-2\right)
\end{equation*}
 of the equation~\eqref{5thKdV} is known~\cite{vaneeva:parkes}.
Using~\eqref{eq_gauge} we get the exact solution of
the equation~\eqref{5thKdVext} with $n=2$,
\begin{equation*}
u=\pm2\sqrt{-10\varepsilon}\left(3\tanh\left(x+24\varepsilon \textstyle{\int}\!e^{-2 \int\!\alpha(t)\,{\rm d}t}{\rm d}t\right)^2-2\right)e^{-\int\! \alpha(t)\, {\rm d}t}.
\end{equation*}

It is worthy to note that the obtained reductions to ODEs can be used for construction of numerical solutions of the generalized fKdV equations, see~\cite{Kuriksha&Posta&Vaneeva2014JPA,VPCS2013} for details.

\section{Conclusion and discussion}
In this paper we present the exhaustive group classification of generalized fKdV equations with time dependent coefficients of the general form~\eqref{eq_gg5kdv}.
The complete result is achieved due to the use of equivalence transformations. We show that up to point equivalence the group classification problem for the initial class can be reduced to a simpler problem for its subclass with $\alpha=0$~\eqref{eq_g5kdv}. After the group classification for the subclass~\eqref{eq_g5kdv} is performed, the most general forms of equations~\eqref{eq_gg5kdv} admitting Lie symmetry extensions can be easily recovered using equivalence transformations. The derived results together with ones obtained in~\cite{KPV_DUBNA} for the case $n=1$ give the complete solution of the group classification problem for nonlinear equations of the form~\eqref{eq_gg5kdv}.

We mentioned in the introduction that Lie symmetry analysis of the class~\eqref{eq_gg5kdv} was initiated in~\cite{Wang&Liu&Zhang2013}, and the case $n=2$ was also treated separately in~\cite{Wang2}. However, the results presented therein are either incorrect~\cite{Wang&Liu&Zhang2013} or incomplete~\cite{Wang2}.
Here we discuss main lucks of the results obtained in those two papers.

In~\cite{Wang2}  only some cases of Lie symmetry
extensions for equations of the form~\eqref{eq_gg5kdv} with $n=2$ were found, namely, the cases with $\alpha={\rm const}$ and $\alpha=1/t$. If one performs the group classification up to the corresponding equivalence transformations it is enough to consider the case $\alpha=0$. If one wants to get the classification, where all equations admitting Lie symmetry extensions are presented, not only their inequivalent representatives, then all such equations will have the coefficient $\alpha$ being arbitrary, so the cases $\alpha={\rm const}$ and $\alpha=1/t$ can  be considered as   particular examples only. Moreover, even studying these particular cases the authors of~\cite{Wang2} missed one case of Lie symmetry extension for each value of $\alpha$ considered by them. For example, for the case $\alpha=0$ this is $\beta=\varepsilon (t+\delta)^\rho,$ where $\varepsilon$, $\delta$ and $\rho$ are arbitrary constants with $\varepsilon\rho\neq0.$ Nevertheless, at least dimensions and basis operators of the found Lie symmetry algebras
 for those particular cases derived in~\cite{Wang2} are correct in contrast to the results presented in~\cite{Wang&Liu&Zhang2013}.

In~\cite{Wang&Liu&Zhang2013} the authors state that they find three cases of Lie symmetry extensions for equations~\eqref{eq_gg5kdv} and in each derived case the corresponding Lie symmetry algebra is four-dimensional. This is a false assertion. In this paper and in~\cite{KPV_DUBNA} we show that equation~\eqref{eq_gg5kdv} admits four-dimensional Lie symmetry algebra if and only if $n=1$ and, moreover, the equation is point-equivalent to the simplest constant-coefficient fKdV equation~$u_t+uu_x+\mu u_{xxxxx}=0,$ where $\mu=\mathrm{const}.$ So, the results of~\cite{Wang&Liu&Zhang2013} are principally incorrect.

In the modern group analysis of differential equations the solution of a group classification problem should be inseparably linked with the study of  admissible transformations in the corresponding class of equations. Neglecting of this often leads to incomplete results as shown in the discussion. Moreover, the knowledge of such transformations can be used for solving other problems concerned with the study of classes of variable-coefficient differential equations or their systems. In particular, in the recent work~\cite{VPS2013} the application of admissible transformations to the study of integrability was analyzed.

\begin{acknowledgement}
The authors would like to thank  the Organizing Committee of LT-10  and especially Prof. Vladimir Dobrev for the hospitality. O.K. and O.V. acknowledge the provided support for their participation in the Workshop.
The authors are also grateful to Vyacheslav Boyko and Roman Popovych for useful discussions.
\end{acknowledgement}

\end{document}